\documentclass[12pt]{article}
\usepackage{amssymb}

\def\hybrid{\topmargin 0pt      \oddsidemargin 0pt
        \headheight 0pt \headsep 0pt
        \voffset=-0.5cm
        \hoffset=-0.25in
        \textwidth 6.75in
        \textheight 9.5in       
        \marginparwidth 0.0in
        \parskip 5pt plus 1pt   \jot = 1.5ex}
\catcode`\@=11
\def\marginnote#1{}

\newcount\hour
\newcount\minute
\newtoks\amorpm
\hour=\time\divide\hour by60 \minute=\time{\multiply\hour by60
\global\advance\minute by-\hour}
\edef\standardtime{{\ifnum\hour<12 \global\amorpm={am}%
        \else\global\amorpm={pm}\advance\hour by-12 \fi
        \ifnum\hour=0 \hour=12 \fi
        \number\hour:\ifnum\minute<10 0\fi\number\minute\the\amorpm}}
\edef\militarytime{\number\hour:\ifnum\minute<10 0\fi\number\minute}

\def\draftlabel#1{{\@bsphack\if@filesw {\let\thepage\relax
   \xdef\@gtempa{\write\@auxout{\string
      \newlabel{#1}{{\@currentlabel}{\thepage}}}}}\@gtempa
   \if@nobreak \ifvmode\nobreak\fi\fi\fi\@esphack}
        \gdef\@eqnlabel{#1}}
\def\@eqnlabel{}
\def\@vacuum{}
\def\draftmarginnote#1{\marginpar{\raggedright\scriptsize\tt#1}}
\def\draftlabel#1{{\@bsphack\if@filesw {\let\thepage\relax
   \xdef\@gtempa{\write\@auxout{\string
      \newlabel{#1}{{\@currentlabel}{\thepage}}}}}\@gtempa
   \if@nobreak \ifvmode\nobreak\fi\fi\fi\@esphack}
        \gdef\@eqnlabel{#1}}
\def\@eqnlabel{}
\def\@vacuum{}
\def\draftmarginnote#1{\marginpar{\raggedright\scriptsize\tt#1}}

\def\draft{\oddsidemargin -.5truein
        \def\@oddfoot{\sl preliminary draft \hfil
        \rm\thepage\hfil\sl\today\quad\militarytime}
        \let\@evenfoot\@oddfoot \overfullrule 3pt
        \let\label=\draftlabel
        \let\marginnote=\draftmarginnote
   \def\@eqnnum{(\theequation)\rlap{\kern\marginparsep\tt\@eqnlabel}%
\global\let\@eqnlabel\@vacuum}  }

\def\numberbysection{\@addtoreset{equation}{section}
        \def\theequation{\thesection.\arabic{equation}}}

\def\underline#1{\relax\ifmmode\@@underline#1\else
        $\@@underline{\hbox{#1}}$\relax\fi}

\def\titlepage{\@restonecolfalse\if@twocolumn\@restonecoltrue\onecolumn
     \else \newpage \fi \thispagestyle{empty}\c@page\z@
        \def\thefootnote{\fnsymbol{footnote}} }

\def\endtitlepage{\if@restonecol\twocolumn \else  \fi
        \def\thefootnote{\arabic{footnote}}
        \setcounter{footnote}{0}}  

\numberbysection \hybrid

\newcounter{mo}

\newcommand{\tr}{{\rm tr}}

\newcommand{\mL}{{\mathcal L}}
\newcommand{\mM}{{\mathcal M}}

\newcommand{\vf}{\varphi}
\newcommand{\al}{\alpha}
\newcommand{\be}{\beta}
\newcommand{\ga}{\gamma}
\newcommand{\om}{\omega}
\newcommand{\vth}{\vartheta}

\newcommand{\Mat}{ {\rm Mat}(N,\mathbb C) }
\newcommand{\MatM}{ {\rm Mat}(M,\mathbb C) }
\newcommand{\MatNM}{ {\rm Mat}(NM,\mathbb C) }

\newcommand{\mC}{\mathbb C}
\newcommand{\mZ}{\mathbb Z}

\newcommand{\mS}{\mathcal S}

\newtheorem{predl}{Proposition}[section]

\def\beq{\begin{equation}}
\def\eq{\end{equation}}
\def\p{\partial}

\def\res{\mathop{\hbox{Res}}\limits}

\begin{document}

\setcounter{page}{1}

\date{}
\date{}
\vspace{30mm}

\begin{flushright}
\end{flushright}
\vspace{0mm}

\begin{center}
 \vspace{0mm} {\LARGE{Relativistic interacting integrable elliptic tops}}
\\
\vspace{14mm} {\large {Andrei Zotov}}\\
 \vspace{10mm}
{\small{\rm Steklov Mathematical Institute of Russian Academy of Sciences,}}\\
{\small{\rm Gubkina str. 8, Moscow, 119991,  Russia}}
\end{center}
%
\begin{center}\footnotesize{{\rm E-mail:}{\rm\ \
 zotov@mi-ras.ru}}\end{center}

 \begin{abstract}
We propose relativistic generalization of integrable systems
describing $M$ interacting elliptic ${\rm gl}(N)$ tops of the
Euler-Arnold type. The obtained models are elliptic integrable
systems, which reproduce the spin elliptic ${\rm GL}(M)$
Ruijsenaars-Schneider model for $N=1$ case, while in the $M=1$ case
they turn into relativistic integrable ${\rm GL}(N)$ elliptic tops.
The Lax pairs with spectral parameter on elliptic curve are
constructed.
 \end{abstract}


{\small{


}}


\section{Introduction}\label{sect1}
\setcounter{equation}{0}

 In \cite{KrichZ} Krichever and Zabrodin suggested the following ansatz for the Lax pair
 with spectral parameter of the spin elliptic ${\rm GL}(M)$
 Ruijsenaars-Schneider model:
  \beq\label{q001}
  \begin{array}{c}
  \displaystyle{
 L_{ij}(z)=S_{ij}\phi(z,q_{ij}+\eta)\,,\quad i,j=1,...,M;
 }
 \\ \ \\
  \displaystyle{
 q_{ij}=q_i-q_j\,,\quad \res\limits_{z=0}L(z)=S\in\MatM\,,
 }
 \end{array}
 \eq
  \beq\label{q002}
  \begin{array}{c}
  \displaystyle{
 M_{ij}(z)=-\delta_{ij}(E_1(z)+E_1(\eta))S_{ii}-(1-\delta_{ij})S_{ij}\phi(z,q_{ij})\,.
 }
 \end{array}
 \eq
 The definitions of the Kronecker function $\phi$ and elliptic functions $E_1$, $E_2$
 are given in the Appendix. Under conditions
  \beq\label{q005}
  \begin{array}{c}
  \displaystyle{
 {S}_{ii}={\dot q}_i\,,\quad i=1,...,M
 }
 \end{array}
 \eq
 the Lax equation
  \beq\label{q0001}
  \begin{array}{c}
  \displaystyle{
 \dot{L}(z)=[L(z),M(z)]
 }
 \end{array}
 \eq
 provides equations of motion for the diagonal part of matrix $S$
  \beq\label{q006}
  \begin{array}{c}
  \displaystyle{
 {\dot S}_{ii}={\ddot q}_i=-\sum\limits_{k:k\neq i}^M
 S_{ik}S_{ki}\Big( E_1(q_{ik}+\eta)+E_1(q_{ik}-\eta)-2E_1(q_{ik}) \Big)
 }
 \end{array}
 \eq
 and for its non-diagonal part (\ref{q0001}) yields
  \beq\label{q007}
  \begin{array}{c}
  \displaystyle{
 {\dot S}_{ij}=\sum\limits_{k:k\neq j}^M
 S_{ik}S_{kj}\Big( E_1(q_{kj}+\eta)-E_1(q_{kj}) \Big)
 -\sum\limits_{k:k\neq i}^M
 S_{ik}S_{kj}\Big( E_1(q_{ik}+\eta)-E_1(q_{ik}) \Big)\,,
 }
  \end{array}
 \eq
or, equivalently,
  \beq\label{q008}
  \begin{array}{c}
  \displaystyle{
 {\dot S}_{ij}
 =S_{ij}(S_{ii}-S_{jj})\Big(E_1(q_{ij}+\eta)-E_1(q_{ij})\Big)+
  }
  \\
    \displaystyle{
 +\sum\limits_{k:k\neq i,j}^M
 S_{ik}S_{kj}\Big( E_1(q_{kj}+\eta)-E_1(q_{kj})-E_1(q_{ik}+\eta)+E_1(q_{ik})
 \Big)\,.
 }
 \end{array}
 \eq
These equations can be viewed as relativistic deformation
\cite{Ruijs} (the deformation parameter is $\eta\in\mathbb{C}$)
  of the spin elliptic Calogero-Moser model \cite{BAB}:
  \beq\label{q0081}
  \begin{array}{c}
  \displaystyle{
 {\ddot q}_i=\sum\limits_{k:k\neq i}^M
 S_{ik}S_{ki}E_2'(q_{ik})\,,
 }
 \\
  \displaystyle{
   {\dot S}_{ii}
 =0\,,\quad
 {\dot S}_{ij}
 =
 \sum\limits_{k:k\neq i,j}^M
 S_{ik}S_{kj}\Big( E_2(q_{ik})-E_2(q_{kj})
 \Big)\,.
 }
 \end{array}
 \eq

The system (\ref{q0081}) admits anisotropic $gl(NM)$ generalization,
where the spin variables $S_{ij}$
 are replaced by matrix-valued variables $\mS^{ij}\in\Mat$.
  Equations of motion are of the form:
  \beq\label{q1537}
  \begin{array}{c}
  \displaystyle{
 {\dot \mS}^{ii}=[\mS^{ii},J(\mS^{ii})]
 +\sum\limits_{k:k\neq i}^M
 \Big( \mS^{ik}J^{q_{ki}}(S^{ki}) - J^{q_{ik}}(\mS^{ik})S^{ki}
 \Big)\,,
 }
 \end{array}
 \eq
  \beq\label{q1538}
  \begin{array}{c}
  \displaystyle{
 {\dot \mS}^{ij}=\mS^{ij}J(\mS^{jj})-J(\mS^{ii})\mS^{ij}+
 \sum\limits_{k:k\neq j}^M \mS^{ik}J^{q_{kj}}(\mS^{kj}) -
 \sum\limits_{k:k\neq i}^M J^{q_{ik}}(\mS^{ik})\mS^{kj}\,,
  }
 \end{array}
 \eq
  \beq\label{q1539}
  \begin{array}{c}
  \displaystyle{
 {\ddot q}_i=-\frac{1}{N}\sum\limits_{k:k\neq i}^M
 \p_{q_i}\tr\Big( J^{q_{ik}}(\mS^{ik})\mS^{ki}
 \Big)\,.
 }
 \end{array}
 \eq
 The anisotropy means the presence (in the equations of motion) of the linear operators $J$, $J^{q_{ij}}$
  acting in the matrix space
 $\Mat$. In the case $N=1$ equations (\ref{q1537})-(\ref{q1539}) reproduce
 (\ref{q0081}), while in the case $M=1$ equation (\ref{q1537}) is simplified to the one of the Euler-Arnold type for the elliptic
 integrable top \cite{LOZ}:
  \beq\label{q1540}
  \begin{array}{c}
  \displaystyle{
  {\dot S}=[S,J(S)]\,,\quad S\in\Mat\,.
 }
 \end{array}
 \eq
 The systems of type (\ref{q1537})-(\ref{q1539}) appeared in papers \cite{Polych} in studies of the matrix models,
 and later they were described as examples of the Hitchin systems
 on ${\rm SL}(NM,\mC)$-bundles
(over elliptic curve) with non-trivial characteristic classes
\cite{LZ,LOSZ2}. Equations
 (\ref{q1537})-(\ref{q1539}) for a more general class of the operators $J$ were obtained
in
 \cite{GSZ}.

 When the matrix of spin variables $\mS=\sum_{ij}E_{ij}\otimes \mS^{ij}\in\MatNM$
 is of rank 1, the r.h.s. of equations (\ref{q1537}) and (\ref{q1539}) are represented in terms of diagonal blocks
of matrix $\mS$ only (i.e. in terms of matrices
 $\mS^{ii}$):
  \beq\label{q2537}
  \begin{array}{c}
  \displaystyle{
 {\dot \mS}^{ii}=[\mS^{ii},J(\mS^{ii})]
 +\sum\limits_{k:k\neq i}^M
 [\mS^{ii},\breve{J}^{q_{ik}}(\mS^{kk})]\,,
  \quad
 {\ddot q}_i=-\frac{1}{N}\sum\limits_{k:k\neq i}^M
 \p_{q_i}\tr\Big( \mS^{ii}\breve{J}^{q_{ik}}(\mS^{kk})
 \Big)\,.
 }
 \end{array}
 \eq
This allows to interpret the equations  as dynamics of $M$
 interacting tops with positions $q_i$. Being written in such a form the model
 resembles the
 initial formulation of the (quantum) spin Calogero-Moser model
 \cite{GH}.

{\bf Purpose of the paper} is to construct relativistic deformation
 of the models (\ref{q1537})-(\ref{q1539}) and (\ref{q2537}).
  We will show that such generalization exists and has the form:
  \beq\label{q1520}
  \begin{array}{c}
  \displaystyle{
 {\dot \mS}^{ii}=[\mS^{ii},J^\eta(\mS^{ii})]
 +\sum\limits_{k:k\neq i}^M
 \Big( \mS^{ik}J^{\eta,\,q_{ki}}(S^{ki}) - J^{\eta,\,q_{ik}}(\mS^{ik})S^{ki}
 \Big)\,,
 }
 \end{array}
 \eq
  \beq\label{q1522}
  \begin{array}{c}
  \displaystyle{
 {\dot \mS}^{ij}=\mS^{ij}J^\eta(\mS^{jj})-J^\eta(\mS^{ii})\mS^{ij}+
 \sum\limits_{k:k\neq j}^M \mS^{ik}J^{\eta,\,q_{kj}}(\mS^{kj}) -
 \sum\limits_{k:k\neq i}^M J^{\eta,\,q_{ik}}(\mS^{ik})\mS^{kj}\,,
  }
 \end{array}
 \eq
  \beq\label{q1510}
  \begin{array}{c}
  \displaystyle{
 {\ddot q}_i=\frac{1}{N}\,\tr\Big({\dot \mS}^{ii}\Big)=\frac{1}{N}\sum\limits_{k:k\neq i}^M
 \tr\Big( \mS^{ik}J^{\eta,\,q_{ki}}(\mS^{ki}) - J^{\eta,\,q_{ik}}(\mS^{ik})\mS^{ki}
 \Big)\,.
 }
 \end{array}
 \eq
 The last equation comes from the trace of both sides of equation
 (\ref{q1520}) together with conditions
  \beq\label{q15101}
  \begin{array}{c}
  \displaystyle{
 {\dot q}_i=\frac{1}{N}\,\tr\Big(\mS^{ii}\Big)\,,\quad
 i=1,...,M\,.
 }
 \end{array}
 \eq
In particular case when the matrix of spin variables is of rank~1
 we get the system of $M$ interacting tops associated with
 $GL(N,\mathbb{C})$ group:
%
  \beq\label{q1720}
  \begin{array}{c}
  \displaystyle{
 {\dot \mS}^{ii}=[\mS^{ii},J^\eta(\mS^{ii})]
 +\sum\limits_{k:k\neq i}^M
 \Big( \mS^{ii}{\widetilde J}^{\eta,\,q_{ki}}(S^{kk}) - {\breve J}^{\eta,\,q_{ik}}(\mS^{kk})S^{ii}
 \Big)\,,
 }
 \end{array}
 \eq
  \beq\label{q17201}
  \begin{array}{c}
  \displaystyle{
 {\ddot q}_i=
 \frac{1}{N}\sum\limits_{k:k\neq i}^M
 \tr\Big( \mS^{ii}{\widetilde J}^{\eta,\,q_{ki}}(S^{kk}) - {\breve J}^{\eta,\,q_{ik}}(\mS^{kk})S^{ii}
 \Big)\,.
 }
 \end{array}
 \eq
 The obtained models can be viewed as anisotropic matrix
 generalization of the spin elliptic ${\rm GL}(M)$ Ruijsenaars-Schneider model, which is reproduced in the $N=1$ case.
  When $M=1$ we get the relativistic deformation of the elliptic top (\ref{q1540}), know previously from
\cite{LOZ8}.

The paper is organized as follows. In Sections~\ref{sect2}
and~\ref{sect3} we give detailed descriptions of the spin elliptic
Ruijsenaars-Schneider
 model and the relativistic elliptic top respectively.
 In Section~\ref{sect4} the model
(\ref{q1520})-(\ref{q1510}) is
 described, and
 $\MatNM$-valued Lax representation with
spectral parameter on elliptic curve is given. In
Section~\ref{sect5}
 we study the case ${\rm rk}(\mS)=1$ and obtain equations (\ref{q1720})-(\ref{q17201}). Explicit
 form of the linear operators $\tilde J^{\eta,q_{ij}}$ and~$\breve J^{\eta,q_{ij}}$
 entering (\ref{q1720}) is derived in the end of Section~\ref{sect5}
 through the usage of the finite-dimensional Fourier transformation of
 elliptic functions. The non-relativistic limit is described as well. It is shown that the limit reproduces
 the previously obtained results \cite{GSZ} â
 in the elliptic case.

\section{Spin Ruijsenaars-Schneider model}\label{sect2}
\setcounter{equation}{0}

In this Section we derive equations of motion. It will help to
simplify the proof of the more complicated statement related to
interacting tops. More precisely, we prove the following
\begin{predl}\label{21}
 Equations of motion
  \beq\label{q009}
  \begin{array}{c}
  \displaystyle{
 {\dot S}_{ii}=-\sum\limits_{k:k\neq i}^M
 S_{ik}S_{ki}\Big( E_1(q_{ik}+\eta)+E_1(q_{ik}-\eta)-2E_1(q_{ik}) \Big)
 }
 \end{array}
 \eq
and (\ref{q007})  are equivalent to the Lax equations with
additional term\footnote{The function~$f$ entering (\ref{q010}) is
defined in (\ref{qq917}), à $\{E_{ij}\}$ -- is the standard basis in
$\MatM$.}:
  \beq\label{q010}
  \begin{array}{c}
  \displaystyle{
 \dot{L}(z)=[L(z),M(z)]+\sum\limits_{i,j=1}^M E_{ij}(\mu_i-\mu_j)S_{ij}f(z,q_{ij}+\eta)
 }
 \end{array}
 \eq
for the pair of matrices  (\ref{q001})-(\ref{q002}) and the set of
variables
  \beq\label{q011}
  \begin{array}{c}
  \displaystyle{
 \mu_i={\dot q}_i-S_{ii}\,,\quad i=1,...,M\,.
 }
 \end{array}
 \eq
 On-shell the constraints  $\mu_i=0$ the matrices (\ref{q001})-(\ref{q002})
 satisfy the Lax equation (\ref{q0001}) and provide equations of motion (\ref{q006})-(\ref{q007}).
\end{predl}

\noindent \underline{\em Proof.}  The additional term is absent in
the diagonal part of (\ref{q010}). Consider the~$i$-th diagonal
element. In the l.h.s. of (\ref{q010}) in $ii$-th element we have
$\dot S_{ii}\phi(z,\eta)$,
 while in the r.h.s. the following expression appears:
  \beq\label{q012}
  \begin{array}{c}
  \displaystyle{
  \sum\limits_{k:k\neq i}L_{ik}M_{ki}-M_{ik}L_{ki}=\sum\limits_{k:k\neq
  i}S_{ik}S_{ki}\Big( \phi(z,q_{ik})\phi(z,q_{ki}+\eta)-\phi(z,q_{ik}+\eta)\phi(z,q_{ki})
  \Big)\,.
 }
 \end{array}
 \eq
 Equation (\ref{q009}) comes from the usage of the relation (\ref{qq915}).
 In the off-diagonal part of the equation (\ref{q010}) for $ij$-th matrix element (with $i\neq j$)
  in the l.h.s. we have
  \beq\label{q013}
  \begin{array}{c}
  \displaystyle{
  {\dot S}_{ij}\phi(z,q_{ij}+\eta)+S_{ij}( {\dot q}_i-{\dot q}_j
  )f(z,q_{ij}+\eta)\,.
 }
 \end{array}
 \eq
And in the r.h.s. we get
  \beq\label{q014}
  \begin{array}{c}
  \displaystyle{
  (M_{jj}-M_{ii})L_{ij}+(L_{ii}-L_{jj})M_{ij}+
 }
  \\ \ \\
  \displaystyle{
  +\sum\limits_{k\neq
  i,j}\Big(L_{ik}M_{kj}-M_{ik}L_{kj}\Big) + (\mu_i-\mu_j)S_{ij}f(z,q_{ij}+\eta)=
 }
 \\ \ \\
  \displaystyle{
  =(S_{ii}-S_{jj})S_{ij}\Big(E_1(z)+E_1(\eta)\Big)\phi(z,q_{ij}+\eta)-(S_{ii}-S_{jj})S_{ij}\phi(z,q_{ij})\phi(z,\eta)-
 }
  \\ \ \\
  \displaystyle{
  -\sum\limits_{k\neq
  i,j}\!S_{ik}S_{kj}\Big( \phi(z,q_{ik}+\eta)\phi(z,q_{kj})-\phi(z,q_{ik})\phi(z,q_{kj}+\eta)
  \Big)+ (\mu_i\!-\!\mu_j)S_{ij}f(z,q_{ij}+\eta)\,.
 }
 \end{array}
 \eq
 Transpose the second term of (\ref{q013}) from the l.h.s. of equation (\ref{q010}) to its r.h.s.
 Then the terms proportional to $(\dot q_i-\dot q_j)S_{ij}$ are
 cancelled out.
 For the terms proportional to $(S_{ii}-S_{jj})S_{ij}$, we get a common factor:
  \beq\label{q015}
  \begin{array}{c}
  \displaystyle{
\Big(E_1(z)+E_1(\eta)\Big)\phi(z,q_{ij}+\eta)-\phi(z,q_{ij})\phi(z,\eta)-f(z,q_{ij}+\eta)\stackrel{(\ref{qq915}),(\ref{qq917})}{=}
 }
  \\ \ \\
  \displaystyle{
=\phi(z,q_{ij}+\eta)\Big(E_1(q_{ij}+\eta)-E_1(q_{ij})\Big)\,.
 }
 \end{array}
 \eq
 Using also relation (\ref{qq915}) for the expression in the sum in
 (\ref{q014}),
 we finally get the off-diagonal part of equations of motion in the
 form (\ref{q008}). It is easily seen that
 the latter is equivalent to (\ref{q007}). This finishes the proof. $\blacksquare$

 The non-relativistic limit appears as follows. Let us redefine the
 time variable
  \beq\label{q017}
  \begin{array}{c}
  \displaystyle{
  t\rightarrow t/\eta\,.
 }
 \end{array}
 \eq
 In particular, it means that $\dot q_i\mapsto\eta\dot q_i$ and~$\ddot q_i\mapsto\eta^2\ddot q_i$.
 From the definition (\ref{qq912}) near $\eta=0$ we have:
  \beq\label{q018}
  \begin{array}{c}
  \displaystyle{
  E_1(q+\eta)=E_1(q)-\eta E_2(q)-\frac{1}{2}\,\eta^2E_2'(q)+O(\eta^3)\,.
 }
 \end{array}
 \eq
 In the limit $\eta\rightarrow 0$ the constraints (\ref{q005}) turn
 into the set of conditions
  \beq\label{q019}
  \begin{array}{c}
  \displaystyle{
  S_{ii}=0\,,\quad i=1,...,M\,,
 }
 \end{array}
 \eq
  and the equations of motion (\ref{q006})-(\ref{q008}) in view of (\ref{q019})
 take the form
  \beq\label{q020}
  \begin{array}{c}
  \displaystyle{
 {\ddot q}_i=\sum\limits_{k:k\neq i}^M
 S_{ik}S_{ki}E_2'(q_{ik})\,.
 }
 \end{array}
 \eq
 Equations of motion for the diagonal part of spin variables are $\dot S_{ii}=0$, and for $i\neq j$
 we have
  \beq\label{q021}
  \begin{array}{c}
  \displaystyle{
 {\dot S}_{ij}
 =
 \sum\limits_{k:k\neq i,j}^M
 S_{ik}S_{kj}\Big( E_2(q_{ik})-E_2(q_{kj})
 \Big)\,.
 }
 \end{array}
 \eq
 In this way we get the equations of motion of the classical spin
Calogero-Moser model \cite{BAB}. Let us remark that the choice of
the constraints   $\mu_i=0$ is not necessary. All derivations are
also valid for the constraints
 $\mu_i=\nu=\hbox{const}$ for all $i=1,\ldots,M$. This is a set of the first class constraints in the Calogero-Moser model.
  They should be supplied with  $M$ conditions of gauge fixation with
  respect to the coadjoint action of the Cartan subgroup of ${\rm
 GL}(M,\mC)$, i.e. with respect to conjugation by diagonal
 matrices. Then the total set  of
 $2M$~conditions forms the second class constraints, and one should perform the Poisson reduction with respect to these constraints.
  The reduction procedure changes equations of motion due to
  reducing the number of independent variables and due to the Dirac terms appearing in the reduced Poisson brackets.

 From all has been said it follows that the equations
 (\ref{q020})-(\ref{q021}) should be considered as
 intermediate stage of the Poisson reduction corresponding to simple
 restriction of the unreduced system (with linear Poisson-Lie
 brackets) to the imposed constraints $\mu_i=\nu$, but the reduction procedure is not performed yet.
 The equations of motion in the relativistic case
 (\ref{q006})-(\ref{q008})  should be understood in the same manner on the
  constraints (\ref{q005}). It should be mentioned that the Poisson
  structure (and the classical $r$-matrix structure) for the spin elliptic Ruijsenaars-Schneider model is
  unknown yet. At the same time for the trigonometric and rational models
  the Poisson structures and the group-theoretical description
  are known
  \cite{AF,Resh,Feher,ChF,AO,PZ}.


\section{Relativistic integrable top}\label{sect3}
\setcounter{equation}{0}
A special basis  in the space $\Mat$ (the sine-algebra basis) is
used for description of elliptic tops.  The basis is of the form
 \beq\label{qq901}
 \begin{array}{c}
  \displaystyle{
 T_\al=T_{\al_1 \al_2}=\exp\left(\frac{\pi\imath}{{ N}}\,\al_1
 \al_2\right)Q^{\al_1}\Lambda^{\al_2}\,,\quad
 \al=(\al_1,\al_2)\in\mZ_{ N}\times\mZ_{ N}\,,
 }
 \end{array}
 \eq
where $Q$ and~$\Lambda$~-- pair of matrices (for which
$Q^N=\Lambda^N=1_{N\times N}$) with elements
 \beq\label{qq902}
 \begin{array}{c}
  \displaystyle{
Q_{kl}=\delta_{kl}\exp(\frac{2\pi
 \imath}{{ N}}k)\,,\ \ \
 \Lambda_{kl}=\delta_{k-l+1=0\,{\hbox{\tiny{mod}}}\,
 { N}}\,,\quad k,l=1,...,N\,.
 }
 \end{array}
 \eq
It is the finite-dimensional representation of the Heisenberg
 group:
 \beq\label{qq903}
 \begin{array}{c}
  \displaystyle{
 \Lambda^{a_2} Q^{a_1}=\exp\left(\frac{2\pi\imath}{{ N}}\,a_1
 a_2\right)Q^{a_1}\Lambda^{a_2}\,,\qquad a_1,a_2\in\mZ\,.
 }
 \end{array}
 \eq
 It is easy to see that the product of basis matrices can be written
 in the form
  \beq\label{qq904}
 \begin{array}{c}
  \displaystyle{
T_\al T_\be=\kappa_{\al,\be} T_{\al+\be}\,,\ \ \
\kappa_{\al,\be}=\exp\left(\frac{\pi \imath}{{ N}}(\be_1
\al_2-\be_2\al_1)\right)\,,
 }
 \end{array}
 \eq
 where $\al+\be=(\al_1+\be_1,\al_2+\be_2)$.
 In
 particular, it follows from the latter that
  \beq\label{q905}
 \begin{array}{c}
  \displaystyle{
\tr(T_\al T_\be)={ N}\delta_{\al,-\be}\,.
 }
 \end{array}
 \eq
The term ``sine-algebra'' comes from the form of the structure
constants of ${\rm gl}(N)$ Lie algebra, which are given as follows
  \beq\label{qq906}
 \begin{array}{c}
  \displaystyle{
[T_\al, T_\be]=C_{\al\be} T_{\al+\be}\,,\quad
C_{\al\be}=\kappa_{\al,\be}-\kappa_{\be,\al}=
2\imath\sin\left(\frac{\pi \imath}{{ N}}(\be_1
\al_2-\be_2\al_1)\right)\,.
 }
 \end{array}
 \eq
 In what follows we denote the index $(0,0)$ as zero for brevity,
 i.e.
  \beq\label{qq907}
 \begin{array}{c}
  \displaystyle{
 T_{(0,0)}=1_N=T_0\,.
 }
 \end{array}
 \eq

\paragraph{Relativistic top} in ${\rm GL}(2)$ case was, in fact, described
 by Sklyanin in \cite{Skl1} as Hamiltonian system with the quadratic Poisson bracket
(the classical Sklyanin algebra)
 through the quasi-classical limit of the quantum exchange (or $RLL$-) relations. Here we use the description of
  ${\rm GL}(N)$ top,
 proposed in papers \cite{LOZ8,LOZ16}.

 Dynamical variables are the components~$S_\alpha$ of the matrix
$S\in\Mat$ in the basis (\ref{qq901}). Let us define the action of
the linear operator
 (it is the multi-dimensional analogue of the inverse inertia tensor in principle axes):
  \beq\label{q201}
 \begin{array}{c}
  \displaystyle{
J^\eta(S)=\sum\limits_{\al\neq 0}T_\al S_\al
J_\al^\eta=\sum\limits_{\al\neq 0}T_\al S_\al\Big(
E_1(\om_\al+\eta)-E_1(\om_\al) \Big)\,,\quad
   S=\sum\limits_{\al}T_\al
S_\al\,,
 }
 \end{array}
 \eq
  where $\om_\al$ is defined as in (\ref{qq913}).
 Equations of
 motion are the Euler-Arnold equations for the dynamics of the rigid
 body in multi-dimensional space:
  \beq\label{q202}
 \begin{array}{c}
  \displaystyle{
 {\dot S}=[S,J^\eta(S)]\,.
 }
 \end{array}
 \eq
 The Lax pair is of the form:
 \beq\label{q203}
 \begin{array}{c}
  \displaystyle{
  L(z)={\sum}_\al T_\al S_\al\vf_\al(z,\eta+\om_\al)\,,\quad\quad
  M(z)=-\sum\limits_{\al\neq 0} T_\al S_\al\vf_\al(z,\om_\al)\,.
 }
 \end{array}
 \eq
 \begin{predl}\label{31}
  The Lax equation (\ref{q0001}) for the pair of matrices (\ref{q203})
 is equivalent to equation of motion (\ref{q202}).
 \end{predl}
\noindent \underline{\em Proof.} The l.h.s. of the Lax equation is
equal to ${\sum}_\al T_\al {\dot S}_\al\vf_\al(z,\eta+\om_\al)$. For
the r.h.s. we have
 \beq\label{q204}
 \begin{array}{c}
  \displaystyle{
 -\sum\limits_{\be,\ga\neq 0} [T_\be,T_\ga] S_\be S_\ga
 \vf_\be(z,\om_\be+\eta)\vf_\ga(z,\om_\ga)\,,
 }
 \end{array}
 \eq
where the term with~$\beta=0$ is absent due to (\ref{qq907}): it is
proportional to the identity matrix.
 Antisymmetrizing this expression with respect to indices~$\beta$
 and~$\gamma$ we get
 \beq\label{q205}
 \begin{array}{l}
  \displaystyle{
  -\frac12\sum\limits_{\be,\ga\neq 0} [T_\be,T_\ga] S_\be S_\ga
  \Big(
  \vf_\be(z,\om_\be+\eta)\vf_\ga(z,\om_\ga)-\vf_\ga(z,\om_\ga+\eta)\vf_\be(z,\om_\be)\Big)\stackrel{(\ref{qq919}),(\ref{qq906})}{=}
 }
 \\ \ \\
   \displaystyle{
  =-\frac12\sum\limits_{\be,\ga\neq 0} C_{\be,\ga} T_{\be+\ga} S_\be S_\ga
  \vf_{\be+\ga}(z,\eta+\om_{\be+\ga})\Big( J_\be^\eta-J_\ga^\eta  \Big)=
 }
  \\ \ \\
   \displaystyle{
  =\sum\limits_{\be,\ga\neq 0} C_{\be,\ga} T_{\be+\ga} S_\be S_\ga
  \vf_{\be+\ga}(z,\eta+\om_{\be+\ga})J_\ga^\eta\,.
 }
 \end{array}
 \eq
 In the basis components (\ref{qq901}) the equations of motion take
 the form
 \beq\label{q206}
 \begin{array}{c}
  \displaystyle{
  {\dot S}_{0}=0\,,\quad\quad {\dot S_\al}=\sum\limits_{\be\neq 0} C_{\be,\al-\be}
  S_\be S_{\al-\be} J_{\al-\be}^\eta\,,\quad \al\neq 0\,,
 }
 \end{array}
 \eq
  which means that the Lax equation is equivalent to
(\ref{q202}). $\blacksquare$

 The non-relativistic limit $\eta\rightarrow 0$ is taken together
 with the rescaling of the time variable $t\rightarrow t/\eta$. Equation of motion
(\ref{q202}) turns into
 Euler-Arnold equation of the non-relativistic elliptic top \cite{LOZ}:
  \beq\label{q207}
 \begin{array}{c}
  \displaystyle{
 {\dot S}=[S,J(S)]\,,\quad
 J(S)=-\sum\limits_{\al\neq 0}T_\al S_\al E_2(\om_\al)\,,\quad
   S=\sum\limits_{\al}T_\al
S_\al\in\Mat\,,
 }
 \end{array}
 \eq
 where the function $E_2$ (\ref{qq912}) is used.

\section{${\rm GL}(NM)$ generalization of spin $M$-body \\ Ruijsenaars-Schneider model}\label{sect4}
\setcounter{equation}{0}
In this Section we define a generalization of the spin
Ruijsenaars-Schneider model from Section~\ref{sect2} and
simultaneously for the relativistic top from Section~\ref{sect3}.
 The Lax representation for new model is of size $NM\times NM$.
 It possesses natural block-matrix structure:
 \beq\label{q501}
 \mL(z)= \left.\left(\begin{array}{cccc}
  \mL^{11}(z)  &  \mL^{12}(z) &  \ldots & \mL^{1M}(z)
  \\ \ \\
 \mL^{21}(z)  & \mL^{22}(z)    & \ldots & \mL^{2M}(z)
 \\
 \vdots & \vdots &  \ddots & \vdots \\
 \mL^{M1}(z)  & \mL^{M2}(z)  & \ldots & \mL^{MM}(z)
 \end{array}\right)\ \right\}
 \
\begin{array}{c}
  \hbox{in each column}
  \\
  M\ \hbox{blocks}
  \\
  \hbox{of size}\ N\times N
\end{array}
 \eq
Put it differently,
 \beq\label{q502}
 \begin{array}{c}
  \displaystyle{
  \mL(z)=\sum\limits_{i,j=1}^M E_{ij}\otimes
  \mL^{ij}(z)\in\MatNM\,,\quad  \mL^{ij}(z)\in\Mat\,.
 }
 \end{array}
 \eq
 Likewise the residue of the Lax matrix (\ref{q001}) was equal to the matrix~$S$, here the residue of the Lax matrix (\ref{q501})
  is equal to
 $\res\limits_{z=0}\mL(z)=\mS\in\MatNM$, and for each block we have
 \beq\label{q5021}
 \begin{array}{c}
  \displaystyle{
 \mL^{ij}(z)=\mL^{ij}(\mS^{ij},z)\,,\quad
 \mS^{ij}=\res\limits_{z=0}\mL^{ij}(z)\in\Mat\,.
 }
 \end{array}
 \eq
  Explicit expression for the $ij$-th block of the Lax matrix takes
 the form
 \beq\label{q503}
 \begin{array}{c}
  \displaystyle{
  \mL^{ij}(z)=\sum\limits_{\al}T_\al\mS^{ij}_\al\vf_\al(z,\om_\al+q_{ij}+\eta)\,,\quad
  q_{ij}=q_i-q_j\,, \quad
 \om_\al=\frac{\al_1+\al_2\tau}{N}\,,
 }
 \end{array}
 \eq
 and for $ij$-th block of the accompany $M$-matrix we get
 \beq\label{q504}
 \begin{array}{c}
  \displaystyle{
  \mM^{ij}(z)=-\sum\limits_{\al}T_\al\mS^{ij}_\al\vf_\al(z,\om_\al+q_{ij})\,,\quad
  \hbox{for}\ i\neq j\,,
 }
 \\
   \displaystyle{
  \mM^{ii}(z)=-T_0\mS^{ii}_0\Big(E_1(z)+E_1(\eta)\Big)-\sum\limits_{\al\neq 0}T_\al\mS^{ii}_\al\vf_\al(z,\om_\al)\,.
 }
 \end{array}
 \eq
 In the case $N=1$ we are left with the scalar parts of the above matrices only, i.e. with those proportional
  to $T_0=1_N$.
 In this way the Lax pair
(\ref{q001})-(\ref{q002}) is reproduced. When $M=1$ we are left with
a single (diagonal) block.
 In this case we get the Lax pair (\ref{q203}) up to the term $-E_1(\eta)T_{0}S^{11}_0$ in the $M$-matrix.
 But for $M=1$ it does not effect the equations of motion since it is proportional to the identity matrix.

 To have a compact form of equations of motion let us define the
following linear operator,
 acting on $ij$-th block (when $i\neq j$) of the matrix~$\mathcal S$:
  \beq\label{q507}
 \begin{array}{c}
  \displaystyle{
J^{\eta,\,q_{ij}}(\mS^{ij})=\sum\limits_{\al}T_\al \mS^{ij}_\al\Big(
E_1(\om_\al+q_{ij}+\eta)-E_1(\om_\al+q_{ij}) \Big)\,.
 }
 \end{array}
 \eq
 At the same time for the diagonal blocks we use the linear operator $J^\eta$  (\ref{q201}).
  As will be shown below, the equations of motion coming from the Lax
 equation with the Lax pair
(\ref{q503})-(\ref{q504}) take the following form for the diagonal
blocks of the matrix  $\mathcal S$:
  \beq\label{q520}
  \begin{array}{c}
  \displaystyle{
 {\dot \mS}^{ii}=[\mS^{ii},J^\eta(\mS^{ii})]
 +\sum\limits_{k:k\neq i}^M
 \Big( \mS^{ik}J^{\eta,\,q_{ki}}(S^{ki}) - J^{\eta,\,q_{ik}}(\mS^{ik})S^{ki}
 \Big)\,,
 }
 \end{array}
 \eq
 and for the non-diagonal blocks we get
  \beq\label{q521}
  \begin{array}{c}
  \displaystyle{
 {\dot \mS}^{ij}=\mS^{ij}J^\eta(\mS^{jj})-J^\eta(\mS^{ii})\mS^{ij}
                 +\mS^{ii}J^{\eta,\,q_{ij}}(\mS^{ij})-J^{\eta,\,q_{ij}}(\mS^{ij})\mS^{jj}+
  }
\\ \ \\
  \displaystyle{
 +\sum\limits_{k:k\neq i,j}^M
 \Big( \mS^{ik}J^{\eta,\,q_{kj}}(\mS^{kj}) - J^{\eta,\,q_{ik}}(\mS^{ik})\mS^{kj}
 \Big)\,,
 }
 \end{array}
 \eq
 or, equivalently,
  \beq\label{q522}
  \begin{array}{c}
  \displaystyle{
 {\dot \mS}^{ij}=\mS^{ij}J^\eta(\mS^{jj})-J^\eta(\mS^{ii})\mS^{ij}+
 \sum\limits_{k:k\neq j}^M \mS^{ik}J^{\eta,\,q_{kj}}(\mS^{kj}) -
 \sum\limits_{k:k\neq i}^M J^{\eta,\,q_{ik}}(\mS^{ik})\mS^{kj}\,.
  }
 \end{array}
 \eq
 For the positions of particles the following equation holds
  \beq\label{q510}
  \begin{array}{c}
  \displaystyle{
 {\ddot q}_i=\frac{1}{N}\,\tr\Big({\dot \mS}^{ii}\Big)=\frac{1}{N}\sum\limits_{k:k\neq i}^M
 \tr\Big( \mS^{ik}J^{\eta,\,q_{ki}}(\mS^{ki}) - J^{\eta,\,q_{ik}}(\mS^{ik})\mS^{ki}
 \Big)\,,
 }
 \end{array}
 \eq
 which is deduced  from equation
 (\ref{q520}) by taking the trace
 of both parts together with the conditions
  \beq\label{q5101}
  \begin{array}{c}
  \displaystyle{
 {\dot q}_i=\mS^{ii}_0=\frac{1}{N}\,\tr\Big(\mS^{ii}\Big)\,,\quad
 i=1,...,M\,,
 }
 \end{array}
 \eq
 -- analogues of relations  (\ref{q005}) for the generalized model.

Notice that for $N=1$ the linear operators~$J^{\eta,q_{ij}}(\mathcal
S^{ij})$ (\ref{q507}) and $J^\eta(\mS^{ii})$ (\ref{q201}) take the
form
  \beq\label{q5071}
 \begin{array}{c}
  \displaystyle{
 N=1:\quad
 J^{\eta,\,q_{ij}}(S_{ij})=(E_1(q_{ij}+\eta)-E_1(q_{ij}))S_{ij}\,,\quad
 J^\eta(S_{ii})=0\,.
 }
 \end{array}
 \eq
The operator~$J^\eta$ is equal to zero in the case $N=1$ due to the
 absence of the scalar term
  (with~$\alpha=0$) in the r.h.s. of equation (\ref{q201}). Using
(\ref{q5071}), it is easy to see that for $N=1$ the equations of
motion (\ref{q520})-(\ref{q522}) and (\ref{q510}) turn into
(\ref{q006})-(\ref{q008}) for $M$-body spin Ruijsenaars-Schneider
model. Let us prove the statement similar to the Proposition
\ref{21}.

\begin{predl}\label{41}
 Equations of motion $\mS$ (\ref{q520}), (\ref{q521})-(\ref{q522})
 for diagonal and off-diagonal
 blocks of matrix ~$\mathcal S$ are equivalent to the Lax equation
 with additional term
  \beq\label{q523}
  \begin{array}{c}
  \displaystyle{
 \dot{\mL}(z)=[\mL(z),\mM(z)]+\sum\limits_{i,j=1}^M \sum\limits_\al
 E_{ij}\otimes T_\al (\mu^i_0-\mu^j_0)\mS^{ij}_\al f_\al(z,\om_\al+q_{ij}+\eta)
 }
 \end{array}
 \eq
 for the Lax pair (\ref{q503})-(\ref{q504}) and the set of  variables
  \beq\label{q524}
  \begin{array}{c}
  \displaystyle{
 \mu^i_0={\dot q}_i-\mS^{ii}_0={\dot q}_i-\frac{1}{N}\,\tr\Big(\mS^{ii}\Big)\,,\quad i=1,...,M\,.
 }
 \end{array}
 \eq
 On-shell constraints $\mu^i_0=0$ the matrices (\ref{q503})-(\ref{q504})
 satisfy the Lax equation $\dot{\mL}(z)=[\mL(z),\mM(z)]$ and provide equations of motion (\ref{q520})-(\ref{q522}) and (\ref{q510}).
\end{predl}
\noindent \underline{\em Proof.} The proof is similar to the one for
Proposition \ref{21}. Consider the l.h.s. of (\ref{q523}):
  \beq\label{q527}
  \begin{array}{c}
  \displaystyle{
 \dot{\mL^{ii}}(z)=\sum\limits_{\al}T_\al\dot{\mS}^{ii}_\al\vf_\al(z,\om_\al+\eta)\,,
 }
 \end{array}
 \eq
For $i\neq j$ we have
 $$
  \displaystyle{
 i\neq j:\quad
 \dot{\mL^{ij}}(z)=\sum\limits_{\al}T_\al
 \Big(\dot{\mS}^{ij}_\al\vf_\al(z,\om_\al+q_{ij}+\eta)+
 ( {\dot q}_i-{\dot q}_j ){\mS}^{ij}_\al
 f_\al(z,\om_\al+q_{ij}+\eta)\Big)=
 }
 $$
  \beq\label{q528}
  \begin{array}{l}
  \displaystyle{
 =\sum\limits_{\al}T_\al\dot{\mS}^{ij}_\al\vf_\al(z,\om_\al++q_{ij}+\eta)+
 \sum\limits_{\al}T_\al(\mu^i_0-\mu^j_0)\mS^{ij}_\al f_\al(z,\om_\al+q_{ij}+\eta)+
  }
  \\ \ \\
    \displaystyle{
 +\sum\limits_{\al}T_\al(\mS^{ii}_0-\mS^{jj}_0){\mS}^{ij}_\al\vf_\al(z,\om_\al+q_{ij}+\eta)
 \Big( E_1(z\!+\!\om_\al\!+\!q_{ij}\!+\!\eta)-E_1(\om_\al\!+\!q_{ij}\!+\!\eta) \Big)\,.
  }
 \end{array}
 \eq
 The expression (function) in the last sum in the r.h.s. is just the
 function
 $f_\al(z,\om_\al+q_{ij}+\eta)$ from (\ref{qq918}), being transformed through (\ref{qq917}).
  The second sum in the r.h.s. of (\ref{q528}) exactly cancels out
  with the additional term from the r.h.s. of equation (\ref{q523}).
 The rest of the terms from the r.h.s.
  (i.e. those coming from the commutator)
  are of the following form for the diagonal block
  \beq\label{q529}
  \begin{array}{c}
  \displaystyle{
 [\mL^{ii},\mM^{ii}]
 +\sum\limits_{k: k\neq i} \mL^{ik}\mM^{ki}- \mM^{ik}\mL^{ki}
 }
 \end{array}
 \eq
 and for the off-diagonal $ij$-th block ($i\neq j)$~-- the rest of
 the terms are
  \beq\label{q530}
  \begin{array}{c}
  \displaystyle{
 i\neq j:\quad \mL^{ii}\mM^{ij}-\mM^{ii}\mL^{ij}
 +\mL^{ij}\mM^{jj}-\mM^{ij}\mL^{jj}
 +\sum\limits_{k: k\neq i,j} \mL^{ik}\mM^{kj}- \mM^{ik}\mL^{kj}\,.
 }
 \end{array}
 \eq

The computations for the diagonal blocks are very similar to those
performed in Sections \ref{sect2} and \ref{sect3}. The first term
(commutator) in (\ref{q529}) provides the first term (also a
commutator) in the r.h.s.
 of the equation of motion (\ref{q520}) in the same way as it was made for the relativistic top
 for the equation (\ref{q202}). The sum in
(\ref{q529}) is simplified similarly to (\ref{q012}):
  \beq\label{q5301}
  \begin{array}{c}
  \displaystyle{
 \mL^{ik}\mM^{ki}-\mM^{ik}\mL^{ki}=
 \sum\limits_{\be,\ga}T_\be T_\ga\mS^{ik}_\be\mS^{ki}_\ga\times
 }
 \\
  \displaystyle{
 \times\Big( \vf_\be(z,\om_\be+q_{ik})\vf_\ga(z,\om_\ga+q_{ki}+\eta)
 - \vf_\be(z,\om_\be+q_{ik}+\eta)\vf_\ga(z,\om_\ga+q_{ki}) \Big)\,.
 }
 \end{array}
 \eq
 Applying~ (\ref{qq919}) to the expression in the brackets we get
  \beq\label{q5302}
  \begin{array}{c}
  \displaystyle{
  \sum\limits_{\be,\ga}T_\be T_\ga
 \mS^{ik}_\be\mS^{ki}_\ga
 \vf_{\be+\ga}(z,\om_{\be+\ga}+q_{ii}+\eta)\Big( E_1(\om_\ga+q_{ki}+\eta)-
 }
 \\
  \displaystyle{
 -E_1(\om_\ga+q_{ki})
 -E_1(\om_\be+q_{ik}+\eta)+E_1(\om_\be+q_{ik}) \Big)\,,
 }
 \end{array}
 \eq
 which yields the sum in the r.h.s. of equations of motion (\ref{q520}).
 The scalar (i.e. zero)
 component, which provides
(\ref{q5101}), corresponds to the terms with~$\beta=-\gamma$ due to
relations (\ref{qq904})-(\ref{q905}).

 Consider the off-diagonal block~$ij$. For the first four terms from (\ref{q530})
  let us write down separately all the summands containing the scalar
 (zero) components of the diagonal blocks of matrix~$\mathcal S$ (i.e. the summands
 containing
$\mathcal S^{ii}_0$):
  \beq\label{q531}
  \begin{array}{c}
  \displaystyle{
 (\mS^{ii}_0-\mS^{jj}_0)\sum\limits_{\al}
 T_\al\mS^{ij}_{\al}\Big( (E_1(z)+E_1(\eta))\vf_\al(z,\om_\al+q_{ij}+\eta)-\vf_\al(z,\om_\al+q_{ij})\phi(z,\eta) \Big)
 }
 \\ \ \\
  \displaystyle{
 \stackrel{(\ref{qq919})}=(\mS^{ii}_0-\mS^{jj}_0)\sum\limits_{\al}
 T_\al\mS^{ij}_{\al}\vf_\al(z,\om_\al+q_{ij}+\eta)\Big( E_1(z+\om_\al+q_{ij}+\eta)-E_1(\om_\al+q_{ij})
 \Big)\,.
 }
 \end{array}
 \eq
 There are the same type terms (containing $\mS^{ii}_0$) in the l.h.s. of equation (\ref{q523}).
  They are in the last sum in (\ref{q528}). Transpose them to the
 r.h.s. and sum up with the result (\ref{q531}). Then we get
  \beq\label{q532}
  \begin{array}{c}
  \displaystyle{
 (\mS^{ii}_0-\mS^{jj}_0)\sum\limits_{\al}
 T_\al\mS^{ij}_{\al}\vf_\al(z,\om_\al+q_{ij}+\eta)\Big(
 E_1(\om_\al+q_{ij}+\eta)-E_1(\om_\al+q_{ij})
 \Big)\,.
 }
 \end{array}
 \eq
 Next, write down the rest of the first four terms in
(\ref{q530}), i.e. the terms, which do not contain~$\mathcal
S^{ii}_0$:
  \beq\label{q533}
  \begin{array}{c}
  \displaystyle{
 {\sum\,}' T_\be T_\ga \mS_\be^{ii}\mS^{ij}_\ga
 \Big( \vf_\be(z,\om_\be)\vf_\ga(z,\om_\ga+q_{ij}+\eta) - \vf_\be(z,\om_\be+\eta)\vf_\ga(z,\om_\ga+q_{ij}) \Big)
 }
 \\ \ \\
  \displaystyle{
 -{\sum\,}' T_\ga T_\be \mS_\ga^{ij}\mS^{jj}_\be
 \Big( \vf_\be(z,\om_\be)\vf_\ga(z,\om_\ga+q_{ij}+\eta) - \vf_\be(z,\om_\be+\eta)\vf_\ga(z,\om_\ga+q_{ij}) \Big)
 \,,
 }
 \end{array}
 \eq
 where
the primed sum ${\sum\,}'$  ~-- is the sum over two indices
$\be,\ga\in\mZ_N^{\times 2}$ with condition $\beta\neq 0$ (the terms
with~$\beta=0$ were already accounted in (\ref{q531})). Applying
(\ref{qq919}), we obtain
  \beq\label{q534}
  \begin{array}{l}
  \displaystyle{
 ={\sum\,}' T_\ga T_\be \mS_\ga^{ij}\mS^{jj}_\be \vf_{\be+\ga}(z,\om_{\be+\ga}+q_{ij}+\eta)
 \Big( E_1(\om_\be+\eta)-E_1(\om_\be) \Big)-
 }
 \\ \ \\
  \displaystyle{
 -{\sum\,}' T_\be T_\ga \mS_\be^{ii}\mS^{ij}_\ga\vf_{\be+\ga}(z,\om_{\be+\ga}+q_{ij}+\eta)
 \Big( E_1(\om_\be+\eta)-E_1(\om_\be) \Big)+
  }
  \\ \ \\
  \displaystyle{
 +{\sum\,}' T_\be T_\ga \mS_\be^{ii}\mS^{ij}_\ga\vf_{\be+\ga}(z,\om_{\be+\ga}+q_{ij}+\eta)
 \Big( E_1(\om_\ga+q_{ij}+\eta)-E_1(\om_\ga+q_{ij}) \Big)-
 }
  \\ \ \\
    \displaystyle{
 -{\sum\,}' T_\ga T_\be \mS_\ga^{ij}\mS^{jj}_\be \vf_{\be+\ga}(z,\om_{\be+\ga}+q_{ij}+\eta)
 \Big( E_1(\om_\ga+q_{ij}+\eta)-E_1(\om_\ga+q_{ij}) \Big)\,.
 }
 \end{array}
 \eq
 Thus, in the r.h.s. of the Lax equation we are left with the terms (\ref{q534})
 and
(\ref{q532}). Notice that addition of the term (\ref{q532}) to
(\ref{q534}) is equivalent to removing the
 primes in the last two sums
 in
(\ref{q534}) since the terms (\ref{q532}) correspond to the index
value $\beta=0$ in these sums.
 Therefore, the sum of expressions (\ref{q532}) and
(\ref{q534}) reproduces the first four terms from the r.h.s. of
equation of
 motion (\ref{q521}).

 It remains to determine the contribution to equation of motion
 coming from the last sum in
(\ref{q530}). For this purpose
 consider the difference
  \beq\label{q535}
  \begin{array}{c}
  \displaystyle{
 \mL^{ik}\mM^{kj}-\mM^{ik}\mL^{kj}=
 \sum\limits_{\be,\ga}T_\be T_\ga\mS^{ik}_\be\mS^{kj}_\ga\times
 }
 \\
  \displaystyle{
 \times\Big( \vf_\be(z,\om_\be+q_{ik})\vf_\ga(z,\om_\ga+q_{kj}+\eta)
 - \vf_\be(z,\om_\be+q_{ik}+\eta)\vf_\ga(z,\om_\ga+q_{kj}) \Big)\,.
 }
 \end{array}
 \eq
 Again, applying (\ref{qq919}) to the expression in the brackets, we
 get the answer
  \beq\label{q536}
  \begin{array}{c}
  \displaystyle{
  \sum\limits_{\be,\ga}T_\be T_\ga
 \mS^{ik}_\be\mS^{kj}_\ga
 \vf_{\be+\ga}(z,\om_{\be+\ga}+q_{ij}+\eta)\Big( E_1(\om_\ga+q_{kj}+\eta)-
 }
 \\
  \displaystyle{
 -E_1(\om_\ga+q_{kj})
 -E_1(\om_\be+q_{ik}+\eta)+E_1(\om_\be+q_{ik}) \Big)\,,
 }
 \end{array}
 \eq
 which yields the sum from the r.h.s. of equation of motion
 (\ref{q521}). This finishes the proof. $\blacksquare$

Similarly to (\ref{q017})-(\ref{q021}) and (\ref{q207}) in the
non-relativistic limit $\eta\rightarrow 0$ we have
  \beq\label{q537}
  \begin{array}{c}
  \displaystyle{
 {\dot \mS}^{ii}=[\mS^{ii},J(\mS^{ii})]
 +\sum\limits_{k:k\neq i}^M
 \Big( \mS^{ik}J^{q_{ki}}(S^{ki}) - J^{q_{ik}}(\mS^{ik})S^{ki}
 \Big)\,,
 }
 \end{array}
 \eq
  \beq\label{q538}
  \begin{array}{c}
  \displaystyle{
 {\dot \mS}^{ij}=\mS^{ij}J(\mS^{jj})-J(\mS^{ii})\mS^{ij}+
 \sum\limits_{k:k\neq j}^M \mS^{ik}J^{q_{kj}}(\mS^{kj}) -
 \sum\limits_{k:k\neq i}^M J^{q_{ik}}(\mS^{ik})\mS^{kj}\,,
  }
 \end{array}
 \eq
 and
  \beq\label{q539}
  \begin{array}{c}
  \displaystyle{
 {\ddot q}_i=-\frac{1}{N}\sum\limits_{k:k\neq i}^M
 \p_{q_i}\tr\Big( J^{q_{ik}}(\mS^{ik})\mS^{ki}
 \Big)\,,
 }
 \end{array}
 \eq
 where $J(\mS^{jj})$ are as in (\ref{q207}), and
  \beq\label{q540}
 \begin{array}{c}
  \displaystyle{
J^{q_{ij}}(\mS^{ij})=-\sum\limits_{\al}T_\al \mS^{ij}_\al
E_2(\om_\al+q_{ij})\,.
 }
 \end{array}
 \eq
 The last equation of motion (\ref{q539}) is obtained similarly to
(\ref{q018}). That is for the operator $J^{\eta,\,q_{ij}}(\mS^{ij})$
(\ref{q507}) we use the following expansion near $\eta=0$:
  \beq\label{q5401}
 \begin{array}{c}
  \displaystyle{
J^{\eta,\,q_{ij}}(\mS^{ij})=\eta\,J^{q_{ij}}(\mS^{ij})+\frac{1}{2}\,\eta^2\,\p_{q_i}J^{q_{ij}}(\mS^{ij})+O(\eta^3)\,.
 }
 \end{array}
 \eq
 For $N=1$ equations (\ref{q537})-(\ref{q539}) turn into equations
(\ref{q020})-(\ref{q021}) for the spin Calogero-Moser model.

 The equations (\ref{q537})-(\ref{q540}) were derived in recent paper
\cite{GSZ}\footnote{In the elliptic case in \cite{GSZ} a slightly
different
 normalization factors are used. In particular,
  the positions of particles are divided by~$N$ everywhere.} for a more general case.
  Constraints (\ref{q524}) in the non-relativistic limit take the
  form
  \beq\label{q541}
 \begin{array}{c}
  \displaystyle{
 \frac{1}{N}\,\tr\Big({ \mS}^{ii}\Big)=0\,,\quad i=1,...,M\,.
 }
 \end{array}
 \eq
 The comment made in the end of Section
\ref{sect2}. is applicable to
 these constraints as well. The gauge fixation is performed in this
 case with respect to the coadjoint action of $M$-dimensional
 subgroup in the Cartan subgroup of ${\rm GL}(NM,\mC)$, i.e. with
 respect to conjugation by matrices of the form $D=\sum_k d_kE_{kk}\otimes T_0$.

\section{Interacting relativistic tops}\label{sect5}
\setcounter{equation}{0}
 In this Section we consider the special case of the ${\rm GL}(NM,\mC)$ model
 from the previous Section when the matrix $\mathcal S$ is of rank 1,
 i.e.
 \beq\label{q701}
 \begin{array}{c}
  \displaystyle{
  \mS^{ij}=\xi^i\otimes\rho^j\in\Mat\,,\quad i,j=1,...,M\,,
 }
 \end{array}
 \eq
 or
 \beq\label{q702}
 \begin{array}{c}
  \displaystyle{
  \mS^{ij}_{ab}=\xi^i_a\rho^j_b\,,\quad i,j=1,...,M\,,\quad
  a,b=1,...,N\,,
 }
 \end{array}
 \eq
where $\xi^i$ -- is a set of $M$~vector-columns of hight ~$N$ each,
and~$\rho^i$ -- is a set of $M$~vector-rows of length~$N$ each.

 The purpose of the Section -- is to rewrite the r.h.s. of equations of motion (\ref{q520})
  for the diagonal blocks in terms of the diagonal blocks only. Then for
 $M$ diagonal blocks we will get a closed system of $M$ matrix
 equations of motion.
 A problem in the usage of conditions (\ref{q701})-(\ref{q702}) is that they are written in the standard basis,
 while for the operators $J^{\eta,\,q_{ki}}$ (\ref{q507}) from equations
 (\ref{q520}) the basis $T_\alpha$ (\ref{qq901}) is used.  Let us use the tensor notation to overcome this difficulty.

Consider an operator  $A$ of the form
 \beq\label{q703}
 \begin{array}{c}
  \displaystyle{
  A(S)=\sum\limits_{\al}T_\al S_\al A_\al\in\Mat\,,\quad
  S=\sum\limits_{\al}T_\al S_\al\in\Mat
 }
 \end{array}
 \eq
and introduce notations
 \beq\label{q704}
 \begin{array}{c}
  \displaystyle{
  A_{12}=\sum\limits_{\al}A_\al T_{\al}\otimes T_{-\al}\in\Mat^{\otimes
  2}\,,
 }
 \end{array}
 \eq
 \beq\label{q7041}
 \begin{array}{c}
  \displaystyle{
  {\breve A}_{12}=\sum\limits_{\al}{\breve A}_\al T_{\al}\otimes T_{-\al}\in\Mat^{\otimes
  2}
 }
 \end{array}
 \eq
and
 \beq\label{q705}
 \begin{array}{c}
  \displaystyle{
  {\breve A}_{12}=A_{12}P_{12}\,,
  }
  \\ \ \\
  \displaystyle{
    {\breve
  A}_{21}=\sum\limits_{\al}{\breve A}_\al T_{-\al}\otimes T_{\al}=\sum\limits_{\al}{\breve A}_{-\al} T_{\al}\otimes T_{-\al}
  =P_{12}{\breve A}_{12}P_{12}=P_{12}A_{12}\,,
 }
 \end{array}
 \eq
 where $P_{12}$ -- is the permutation operator\footnote{The set of matrices
$\{e_{ab}\}$ in (\ref{q706}) forms the standard basis of the space
$\Mat$.}
 \beq\label{q706}
 \begin{array}{c}
  \displaystyle{
  P_{12}=\sum\limits_{a,b=1}^N e_{ab}\otimes
  e_{ba}=\frac{1}{N}\,\sum\limits_{\al} T_\al\otimes T_{-\al}\,.
 }
 \end{array}
 \eq
Let us write down a few main properties of the permutation operator:
$(P_{12})^2=1_N\otimes 1_N$ and $(B\otimes C) P_{12}=P_{12}
(C\otimes B)$ for matrices $B,C\in\Mat$.  Also, using the standard
notations (see e.g. \cite{Skl1}) $S_1=S\otimes 1_N$ and $S_2= 1_N
\otimes S$, we have $S_2P_{12}= P_{12}S_1$ and
 \beq\label{q707}
 \begin{array}{c}
  \displaystyle{
  \tr_2(P_{12}S_2)=NS_1\,,
 }
 \end{array}
 \eq
 where $\tr_2$~-- is a trace over the second tensor component.

Using the above mentioned notations the operator (\ref{q703})
  takes the form
 \beq\label{q708}
 \begin{array}{c}
  \displaystyle{
  A(S)\stackrel{(\ref{q905})}=\frac{1}{N}\,\tr_2(A_{12}S_2)=\frac{1}{N}\,\tr_2({\breve
  A}_{12}P_{12}S_2)
  =\frac{1}{N}\,\tr_2(S_2{\breve A}_{12}P_{12})=
 }
 \\ \ \\
  \displaystyle{
 =\frac{1}{N}\,\tr_2(S_2P_{12}{\breve A}_{21})=\frac{1}{N}\,\tr_2(P_{12}S_1{\breve
 A}_{21})\,,
  }
 \end{array}
 \eq
where in the last equality in the first line we used a cyclic
permutation of matrices in the second tensor component.

 Consider expression
 \beq\label{q709}
 \begin{array}{c}
  \displaystyle{
  \mS^{ik}A(\mS^{ki})\stackrel{(\ref{q708})}
  =\frac{1}{N}\,\tr_2\Big( \mS^{ik}_1 P_{12} \mS^{ki}_1{\breve A}_{21}
  \Big)\,.
 }
 \end{array}
 \eq
 For the matrix (\ref{q701}) we get
 \beq\label{q710}
 \begin{array}{c}
  \displaystyle{
   \mS^{ik}_1 P_{12} \mS^{ki}_1=\sum\limits_{a,b=1}^N \Big((\xi^i\otimes\rho^k)e_{ab}(\xi^k\otimes\rho^i)\Big)\otimes
  e_{ba}=\sum\limits_{a,b=1}^N (\xi^i\otimes\rho^i) (\rho^k
  e_{ab}\xi^k)\otimes e_{ba}=
 }
 \\ \ \\
  \displaystyle{
 =\sum\limits_{a,b=1}^N \mS^{ii} \tr(e_{ab}\mS^{kk})\otimes
 e_{ba}=\mS^{ii}_1 \mS^{kk}_2\,,
  }
 \end{array}
 \eq
 where we used that $(\rho^k
  e_{ab}\xi^k)$ is a scalar. Then for (\ref{q709})  we have
 \beq\label{q711}
 \begin{array}{c}
  \displaystyle{
  \mS^{ik}A(\mS^{ki})=\frac{1}{N}\,\mS^{ii} \tr_2\Big( {\breve
  A}_{21}\,\mS^{kk}_2 \Big)\,.
 }
 \end{array}
 \eq
 Similarly,
 \beq\label{q712}
 \begin{array}{c}
  \displaystyle{
  A(\mS^{ik})\mS^{ki}=
  \frac{1}{N}\,\tr_2\Big( {\breve A}_{12} \mS^{ik}_1 P_{12} \mS^{ki}_1 \Big)
  =\frac{1}{N}\, \tr_2\Big( {\breve
  A}_{12}\,\mS^{kk}_2 \Big)\mS^{ii}\,.
 }
 \end{array}
 \eq
 By applying (\ref{q711}) and (\ref{q712}) to equation (\ref{q510}),
we get the following statement.
 \begin{predl}
In the case (\ref{q701}) when the matrix of spin variables is of
rank one the equations of motion (\ref{q520})  take the following
form
  \beq\label{q720}
  \begin{array}{c}
  \displaystyle{
 {\dot \mS}^{ii}=[\mS^{ii},J^\eta(\mS^{ii})]
 +\sum\limits_{k:k\neq i}^M
 \Big( \mS^{ii}{\widetilde J}^{\eta,\,q_{ki}}(S^{kk}) - {\breve J}^{\eta,\,q_{ik}}(\mS^{kk})S^{ii}
 \Big)\,,
 }
 \end{array}
 \eq
  \beq\label{q7201}
  \begin{array}{c}
  \displaystyle{
 {\ddot q}_i=\frac{1}{N}\,\tr\Big({\dot \mS}^{ii}\Big)=
 \frac{1}{N}\sum\limits_{k:k\neq i}^M
 \tr\Big( \mS^{ii}{\widetilde J}^{\eta,\,q_{ki}}(S^{kk}) - {\breve J}^{\eta,\,q_{ik}}(\mS^{kk})S^{ii}
 \Big)\,,
 }
 \end{array}
 \eq
 where\footnote{Explicit expressions for $\tilde J^{\eta,q_{ij}}$, $\breve J^{\eta,q_{ij}}$ (\ref{q721})-(\ref{q722})
  are given below in (\ref{q729})-(\ref{q730}).}
  \beq\label{q721}
 \begin{array}{c}
  \displaystyle{
 {\widetilde J}^{\eta,\,q_{ij}}(\mS^{kk})=
 \frac{1}{N}\, \tr_2\Big({\breve J}_{21}^{\eta,\,q_{ij}}\mS^{kk}_2 \Big)\,,
 }
 \end{array}
 \eq
  \beq\label{q722}
 \begin{array}{c}
  \displaystyle{
 {\breve J}^{\eta,\,q_{ij}}(\mS^{kk})=
 \frac{1}{N}\, \tr_2\Big({\breve J}_{12}^{\eta,\,q_{ij}}\mS^{kk}_2 \Big)
 }
 \end{array}
 \eq
and
  \beq\label{q723}
 \begin{array}{c}
  \displaystyle{
 J_{12}^{\eta,\,q_{ij}}=\sum\limits_{\al}T_\al\otimes T_{-\al}
 \Big(E_1(\om_\al+q_{ij}+\eta)-E_1(\om_\al+q_{ij}) \Big)\,.
 }
 \end{array}
 \eq
 \end{predl}
 Let us derive explicit expression for $\breve J_{12}^{\eta,q_{ij}}=J_{12}^{\eta,q_{ij}}P_{12}$ from
(\ref{q721})-(\ref{q722}):
  \beq\label{q724}
 \begin{array}{c}
  \displaystyle{
 J_{12}^{\eta,\,q_{ij}}P_{12}=\frac1N\sum\limits_{\al,\ga}T_\al T_\ga\otimes
 T_{-\al} T_{-\ga}
 \Big(E_1(\om_\al+q_{ij}+\eta)-E_1(\om_\al+q_{ij}) \Big)\stackrel{(\ref{qq904})}=
 }
 \\ \ \\
  \displaystyle{
 =\frac1N\sum\limits_{\al,\ga}\kappa_{\al,\ga}^2 T_{\al+\ga}\otimes
 T_{-\al-\ga}
 \Big(E_1(\om_\al+q_{ij}+\eta)-E_1(\om_\al+q_{ij}) \Big)=
 }
  \\ \ \\
  \displaystyle{
 =\frac1N\sum\limits_{\ga}T_{\ga}\otimes
 T_{-\ga}\sum\limits_{\al}
 \kappa_{\al,\ga}^2
 \Big(E_1(\om_\al+q_{ij}+\eta)-E_1(\om_\al+q_{ij}) \Big)\,.
 }
 \end{array}
 \eq
 The latter sum over index $\alpha$ is the finite-dimensional Fourier transformation of the expression in the brackets.
 Let us use the formulae (see \cite{Z2}):
  \beq\label{q725}
 \begin{array}{c}
  \displaystyle{
\frac{1}{N}\sum\limits_{\al}
\Big(E_1(\om_\al+\eta)+2\pi\imath\,\p_\tau\om_\al\Big)=E_1(N\eta)
  }
 \end{array}
 \eq
 and
  \beq\label{q726}
 \begin{array}{c}
  \displaystyle{
\frac{1}{N}\sum\limits_{\al} \kappa_{\al,\ga}^2
\Big(E_1(\om_\al+\eta)+2\pi\imath\,\p_\tau\om_\al\Big)=\vf_\ga(N\eta,\om_\ga)\,,\quad\hbox{for}\
\ga\neq 0\,.
  }
 \end{array}
 \eq
 Plugging it into (\ref{q724}), we get
  \beq\label{q727}
 \begin{array}{c}
  \displaystyle{
 {\breve J}_{12}^{\eta,\,q_{ij}}=J_{12}^{\eta,\,q_{ij}}P_{12}=\sum\limits_{\al}
 I_\al^{\eta,\,q_{ij}}\,
  T_{\al}\otimes T_{-\al}\,,
  }
 \end{array}
 \eq
 where
  \beq\label{q728}
 \begin{array}{c}
  \displaystyle{
 I_0^{\eta,\,q_{ij}}=E_1(Nq_{ij}+N\eta)-E_1(Nq_{ij})\,,
 }
 \\ \ \\
  \displaystyle{
 I_\al^{\eta,\,q_{ij}}=\vf_\al(Nq_{ij}+N\eta,\om_\al)-\vf_\al(Nq_{ij},\om_\al)\,,\quad
 \al\neq 0\,.
  }
 \end{array}
 \eq
 In this way we also get explicit answer for
(\ref{q721})-(\ref{q722}):
  \beq\label{q729}
 \begin{array}{c}
  \displaystyle{
 {\breve J}^{\eta,\,q_{ij}}(\mS^{kk})=
 \frac{1}{N}\, \tr_2\Big({\breve J}_{12}^{\eta,\,q_{ij}}\mS^{kk}_2
 \Big)=\sum\limits_{\al} T_\al\mS^{kk}_\al I_\al^{\eta,\,q_{ij}}
 }
 \end{array}
 \eq
 and
  \beq\label{q730}
 \begin{array}{c}
  \displaystyle{
 {\widetilde J}^{\eta,\,q_{ij}}(\mS^{kk})=
 \frac{1}{N}\, \tr_2\Big({\breve J}_{21}^{\eta,\,q_{ij}}\mS^{kk}_2 \Big)=\sum\limits_{\al} T_\al\mS^{kk}_\al
 I_{-\al}^{\eta,\,q_{ij}}\,.
 }
 \end{array}
 \eq
In the non-relativistic limit $\eta\rightarrow 0$ (\ref{q017}) we
have
  \beq\label{q733}
  \begin{array}{c}
  \displaystyle{
 {\dot \mS}^{ii}=[\mS^{ii},J(\mS^{ii})]
 +\sum\limits_{k:k\neq i}^M
 [\mS^{ii},\breve{J}^{q_{ik}}(\mS^{kk})]\,,
  }
  \\
  \displaystyle{
 {\ddot q}_i=-\frac{1}{N}\sum\limits_{k:k\neq i}^M
 \p_{q_i}\tr\Big( \mS^{ii}\breve{J}^{q_{ik}}(\mS^{kk})
 \Big)\,,
 }
 \end{array}
 \eq
  where
  \beq\label{q734}
 \begin{array}{c}
  \displaystyle{
 {\breve J}^{q_{ij}}(\mS^{kk})=\p_\eta\left.{\breve
 J}^{\eta,\,q_{ij}}(\mS^{kk})\right|_{\eta=0}=
 \p_{q_i}\sum\limits_{\al} T_\al\mS^{kk}_\al F_\al^{q_{ij}}
 }
 \end{array}
 \eq
 and
  \beq\label{q735}
 \begin{array}{c}
  \displaystyle{
 F_0^{q_{ij}}=E_1(Nq_{ij})\,,\quad
 F_\al^{q_{ij}}=\vf_\al(Nq_{ij},\om_\al)\,.
 }
 \end{array}
 \eq
 Such answer follows due to evenness of the function  $E_2(z)$. This
 leads to the properties:
  \beq\label{q736}
 \begin{array}{c}
  \displaystyle{
 \p_\eta\left. {\breve J}_{21}^{\eta,\,q}\right|_{\eta=0}
 =\p_\eta\left. {\breve J}_{12}^{\eta,\,-q}\right|_{\eta=0}
 = \p_\eta\left. {\widetilde J}_{12}^{\eta,\,-q}\right|_{\eta=0}
 = \p_\eta\left. {\widetilde J}_{21}^{\eta,\,q}\right|_{\eta=0}\,.
 }
 \end{array}
 \eq

The classical spin variables in the models of Calogero and
Ruijsenaars types are often described in terms of quiver
parametrization \cite{BAB,AF,Polych,ChF}. For ${\rm GL}(M\,,\mC)$
model the latter means introducing $2NM$~variables $\xi^i_a$,
$\rho^i_a$, $i=1,\ldots,M$, $a=1,\ldots,N$,
 so that the spin variables (in ${\rm GL}(M\,,\mC)$ case) are of the form $S_{ij}=\sum_{a}\xi^i_a\rho^j_a$.
 In the trigonometric and rational cases the Poisson structure is known, and one can write equations of motion for the set of
  $\Mat$-valued
 variables~$S^i_{}$, $S^i_{ab}=\xi^i_a\rho^i_b$ (see e.g. the papers \cite{BAB,AF}).
 Such equations can be viewed as isotropic analogues of the equations
(\ref{q537})-(\ref{q539}).  In our approach we deal with
$\MatNM$-valued variable $\mathcal S$, and~$2NM$-dimensional
parametrization is given not by the pair of rectangular matrices of
size
  $N\times M$,
 but rather arises as a particular case (of rank 1) for the matrix of size $NM\times NM$, as in
(\ref{q702}).


\section{Elliptic functions}
\def\theequation{A.\arabic{equation}}
\setcounter{equation}{0} The main tool for construction of the Lax
pairs with spectral parameter on elliptic curve~$\Sigma_\tau$ with
moduli $\tau$  (${\rm Im}(\tau)>0$) is the Kronecker function
  \beq\label{qq911}
  \begin{array}{l}
  \displaystyle{
 \phi(z,q)=\frac{\vth'(0)\vth(z+q)}{\vth(z)\vth(q)}\,,
 }
 \end{array}
 \eq
 defined via the Riemann theta-function
  \beq\label{qq9118}
  \begin{array}{l}
  \displaystyle{
\vth(z)=\displaystyle{\sum _{k\in \mathbb Z}} \exp \left ( \pi
\imath \tau (k+\frac{1}{2})^2 +2\pi \imath
(z+\frac{1}{2})(k+\frac{1}{2})\right )\,,
 }
 \end{array}
 \eq
which has a simple zero at~$z=0$ due to its oddness. Also we use
 the first and the second Eisenstein functions
  \beq\label{qq912}
  \begin{array}{c}
  \displaystyle{
E_1(z)=\frac{\vth'(z)}{\vth(z)}\,,\quad\quad
E_2(z)=-\p_zE_1(z)=\wp(z)-\frac{1}{3}\frac{\vth'''(0)}{\vth'(0)}\,,
 }
 \end{array}
 \eq
 where $\wp(z)$ -- is the Weierstrass  $\wp$-function.
The function $E_2(z)$ is double-periodic on the lattice
$\mathbb{Z}+\tau\mathbb{Z}$ and has the second order pole at $z=0$.
The first Eisenstein function and the Kronecker function have simple
pole at zero with the residue equal to one. They transform on the
lattice as follows:
  \beq\label{qq9121}
  \begin{array}{c}
  \displaystyle{
 E_1(z+1)=E_1(z)\,,\quad\quad E_1(z+\tau)=E_1(z)-2\pi\imath\,,
 }
 \end{array}
 \eq
  \beq\label{qq9122}
  \begin{array}{c}
  \displaystyle{
 \phi(z+1,q)=\phi(z,q)\,,\quad\quad \phi(z+\tau,q)=e^{-2\pi\imath
 q}\phi(z,q)\,.
 }
 \end{array}
 \eq
 The main relation for the function (\ref{qq911}) is the Fay identity of genus one:
  \beq\label{qq914}
  \begin{array}{c}
  \displaystyle{
\phi(z_1,q_1)\phi(z_2,q_2)=\phi(z_1-z_2,q_1)\phi(z_2,q_1+q_2)+\phi(z_2-z_1,q_2)\phi(z_1,q_1+q_2)\,.
 }
 \end{array}
 \eq
 We use its degeneration for derivation of the Lax equation:
  \beq\label{qq915}
  \begin{array}{c}
  \displaystyle{
 \phi(z,q_1)\phi(z,q_2)=\phi(z,q_1+q_2)(E_1(z)+E_1(q_1)+E_1(q_2)-E_1(q_1+q_2+z))\,.
 }
 \end{array}
 \eq
 The following notation is used for the derivative of the Kronecker
 function with respect to the second argument:
 \beq\label{qq917}
 \begin{array}{c}
  \displaystyle{
f(z,q)=\p_q\phi(z,q)\stackrel{(\ref{qq912})}=\phi(z,q)(E_1(z+q)-E_1(q))\,.
 }
 \end{array}
 \eq
 For description of the models of elliptic tops we also define the
 set of function numerated by the index
 $\alpha=(\alpha_1,\alpha_2)\in\mathbb{Z}_N\times\mathbb{Z}_N$ in accordance with
 numeration of elements of the matrix basis (\ref{qq901}):
 \beq\label{qq913}
 \begin{array}{c}
  \displaystyle{
 \vf_\al(z,\om_\al+\eta)=\exp(2\pi\imath\frac{\al_2}{N}\,z)\,\phi(z,\om_\al+\eta)\,,\quad
 \om_\al=\frac{\al_1+\al_2\tau}{N}\,.
 }
 \end{array}
 \eq
Similarly,
 \beq\label{qq918}
 \begin{array}{c}
  \displaystyle{
 f_\al(z,\om_\al+\eta)=\exp(2\pi\imath\frac{\al_2}{N}\,z)\,f(z,\om_\al+\eta)\,.
 }
 \end{array}
 \eq
 Due to (\ref{qq915}) the set of functions (\ref{qq913}) satisfies the following relations:
 \beq\label{qq919}
 \begin{array}{c}
  \displaystyle{
 \vf_\al(z,\om_\al+q_1)\vf_\be(z,\om_\be+q_2)=
 }
 \\ \ \\
  \displaystyle{
=\!\vf_{\al+\be}(z,\om_{\al+\be}\!+\!q_1\!+\!q_2)\Big(
E_1(z)+E_1(\om_\al\!+\!q_1)+E_1(\om_\be\!+\!q_2) -
E_1(z\!+\!\om_{\al+\be}\!+\!q_1\!+\!q_2) \Big)\,.
 }
 \end{array}
 \eq

\subsubsection*{Acknowledgment}
This work is supported by the Russian Science Foundation under grant
19-11-00062.
%


\begin{small}

\end{small}

\end{document}